\documentclass[english]{paper}
\usepackage{lmodern}
\usepackage[T1]{fontenc}
\usepackage[utf8]{inputenc}
\usepackage{geometry}
\usepackage{textcomp}
\usepackage{mathtools}
\usepackage{amsmath}
\usepackage{graphicx}

\makeatletter
\newcommand{\lyxaddress}[1]{
	\par {\raggedright #1
	\vspace{1.4em}
	\noindent\par}
}

\usepackage[unicode=true, bookmarks=true,bookmarksnumbered=false,bookmarksopen=false, breaklinks=false,pdfborder={0 0 0},pdfborderstyle={},backref=false,colorlinks=false]{hyperref}
\hypersetup{pdftitle={Considerations towards quantitative X-ray and neutron tensor tomography: on the validity of linear approximations of dark-field anisotropy}, pdfauthor={Jonas Graetz}}
\usepackage{scalerel}
\usepackage{tikz}
\usetikzlibrary{svg.path}
\definecolor{orcidlogocol}{HTML}{A6CE39}
\tikzset{
  orcidlogo/.pic={
    \fill[orcidlogocol] svg{M256,128c0,70.7-57.3,128-128,128C57.3,256,0,198.7,0,128C0,57.3,57.3,0,128,0C198.7,0,256,57.3,256,128z};
    \fill[white] svg{M86.3,186.2H70.9V79.1h15.4v48.4V186.2z}
                 svg{M108.9,79.1h41.6c39.6,0,57,28.3,57,53.6c0,27.5-21.5,53.6-56.8,53.6h-41.8V79.1z M124.3,172.4h24.5c34.9,0,42.9-26.5,42.9-39.7c0-21.5-13.7-39.7-43.7-39.7h-23.7V172.4z}
                 svg{M88.7,56.8c0,5.5-4.5,10.1-10.1,10.1c-5.6,0-10.1-4.6-10.1-10.1c0-5.6,4.5-10.1,10.1-10.1C84.2,46.7,88.7,51.3,88.7,56.8z};
  }
}
\newcommand\orcidicon[1]{\href{https://orcid.org/#1}{\mbox{\scalerel*{
\begin{tikzpicture}[yscale=-1,transform shape]
\pic{orcidlogo};
\end{tikzpicture}
}{|}}}}

\usepackage{fancyhdr}
\usepackage[font={small,sf},labelfont=bf,format=plain]{caption}

\makeatother

\usepackage{babel}
\begin{document}
\title{Considerations towards quantitative X-ray and neutron tensor tomography:
on the validity of linear approximations of dark-field anisotropy}
\author{Jonas Graetz\,\orcidicon{0000-0002-4403-3686}}
\maketitle

\lyxaddress{\emph{\small{}Universität Würzburg, Lehrstuhl für Röntgenmikroskopie,
Würzburg, Germany}\\
\emph{\small{}Fraunhofer IIS, Magentic Resonance and X-ray Imaging
Department, Würzburg, Germany}}
\begin{abstract}

The validity of two approximative linear tensor models to be used
for grating based X-ray or neutron dark-field tensor tomography is
investigated in a simulation study. While the dark-field contrast
originating from anisotropic microscopic mass distributions has, in
a previous study, been confirmed to be in general a non-linear function
of two orientations (optical axis and axis of interferometer sensitivity),
linear approximations with a reduced parameter space (considering
only one of the orientation dependencies) are highly preferable with
respect to tomographic volume reconstruction from projections. By
regarding isolated volume elements and systematically exploring the
full range of possible anisotropies, direct correspondences are drawn
between the respective tensors characterizing the complete model used
for signal synthesization and the reduced linear models used for reconstruction.
The tensors' dominant orientations are found to agree to a typical
accuracy of 1°, with their eigenspectra exhibiting fuzzy, yet almost
linear relations among each other. Although modeling only either of
two orientation dependencies for the purpose of tensor reconstruction,
the data acquisition scheme must nevertheless adequately address both
dependencies.\thispagestyle{fancy}
\end{abstract}

\section{Introduction}

The anisotropic nature of X-ray and neutron dark-field contrast allows
imaging of directional information within unresolved substructure
of a sample and thereby provides the fundamental prerequisite to tensor-valued
volume imaging analog to other anisotropic contrast modalities e.g.\
in the field of magnetic resonance imaging. Following on demonstrations
of planar dark-field anisotropy e.g.\ by Jensen et al.\ \cite{Jensen2010PRB}, extensions
of anisotropic dark-field imaging to non-scalar volume reconstruction techniques
have been shown by Malecki et al., Bayer et al., Vogel et al., Wieczorek
et al., Dittmann et al., Gao et al., Kim et al. \cite{Malecki2014,Bayer2014,Vogel2015,Wieczorek2016,Dittmann2017,ZGao2019,JKim2020}.
Common to all of the present approaches is a heuristic 3D extension
of planar dark-field anisotropy as a function of the interferometer
sensitivity axis orientation, while the specific signal models and
reconstruction algorithms vary considerably.  Although formal extensions
of the Radon transform and its inverse to vector and tensor fields
(c.f.\ e.g.\ the overview given by Defrise and Gullberg \cite{Defrise2005})
establish the theoretic feasibility of non-scalar tomography, the
linear projection models assumed and required within the mathematical
conception of tensor tomography are not actually well reproduced by
the available physical contrast modality (dark-field). 

In a recent review, the origination and actual orientation dependence
of dark-field contrast for anisotropic mass distributions has been
investigated in detail (Graetz et al.\ \cite{Graetz2020}), yielding
a minimal yet non-linear model capturing the central features of general
dark-field anisotropy including in particular also the effect of varying
scattering cross section in addition to the characteristic dependence
of dark-field contrast on a structure's correlation lengths. I.e.,
general dark-field anisotropy is in particular concluded to be a function
of two orientations (the optical axis and a perpendicular axis of
interferometer sensitivity). 

Although this additional complexity, and the non-linearity in particular,
are highly undesirable with regard to tomographic reconstruction,
the derived model allows to synthesize large amounts of anisotropic
dark-field signals that will here be used to systematically study
the applicability of approximative linear tensor models amenable to
classic tensor tomography. As it likewise uses a tensor to parametrize
the structural anisotropy of the considered volume element, direct
comparisons between the respective tensors of the physically motivated
non-linear signal model and the mathematically motivated linear tensor
models can be made. The analyses form the basis towards a more detailed
and quantitative understanding of X-ray or neutron dark-field tensor
tomography. 

\section{Methods}

\subsection{Physical model of dark-field anisotropy}

The following minimal model for arbitrarily oriented anisotropic mass
density distributions, modeled as Gaussian ellipsoids, has been motivated
in Graetz et al.\ 2020 \cite{Graetz2020}:
\begin{align}
\mu_{\mathrm{DF}}(\boldsymbol{T}) & =-\ln(v)\propto\frac{1}{\sqrt{T_{\mathrm{zz}}}}(T_{\mathrm{xx}}-\frac{T_{\mathrm{xz}}^{2}}{T_{\mathrm{zz}}})\label{eq:dark-field}\\
\text{for }\hat{n} & =(0,0,1)\nonumber \\
\text{and }\hat{e} & =(1,0,0)\nonumber \\
\text{with }\boldsymbol{T} & =\left[\begin{array}{ccc}
T_{\mathrm{xx}} & T_{\mathrm{xy}} & T_{\mathrm{xz}}\\
T_{\mathrm{xy}} & T_{\mathrm{xx}} & T_{\mathrm{yz}}\\
T_{\mathrm{xz}} & T_{\mathrm{yz}} & T_{\mathrm{zz}}
\end{array}\right]=\boldsymbol{R}\left[\begin{array}{ccc}
\sigma_{1}^{-2} & 0 & 0\\
0 & \sigma_{2}^{-2} & 0\\
0 & 0 & \sigma_{3}^{-2}
\end{array}\right]\boldsymbol{R}^{T}\nonumber 
\end{align}
where $\mu_{\mathrm{DF}}$ denotes the dark-field contrast, $\sigma_{i}$
the ellipsoid's standard deviations along its principal axes and $\boldsymbol{R}$
characterizes its orientation relative to the optical path $\hat{n}$
and the axis of grating sensitivity $\hat{e}$. $v\in[0,1]$ is the
sample-induced visibility loss extracted from grating interferometric
images. General orientations of the acquisition system characterized
by $\hat{n}$ and $\hat{e}$ other than $(0,0,1)$ and $(1,0,0)$
can be considered by a respective additional (inverse) rotation transformation
applied to $\boldsymbol{T}$ prior to evaluating Eq.~\ref{eq:dark-field}.

The given model derives analytically from the wave optical approach
to the theoretical description of dark-field contrast given by Yashiro
et al.\ and Lynch et al.\ \cite{Yashiro2010,Lynch2011} and has
been experimentally confirmed to reproduce, despite a number of first
order approximations, the characteristic signal dependencies on inclinations
both with respect to the optical axis and with respect to the grating
interferometer's axis of sensitivity (\cite{Graetz2020}).

Eq.~\ref{eq:dark-field} will be used here to synthesize dark-field
signals required for the systematic evaluation of tensor reconstruction
techniques based on the mathematically motivated linear signal models
addressed in the following Section.

\begin{figure}
\begin{centering}
\includegraphics[viewport=0bp 0bp 511bp 273bp,width=0.45\columnwidth]{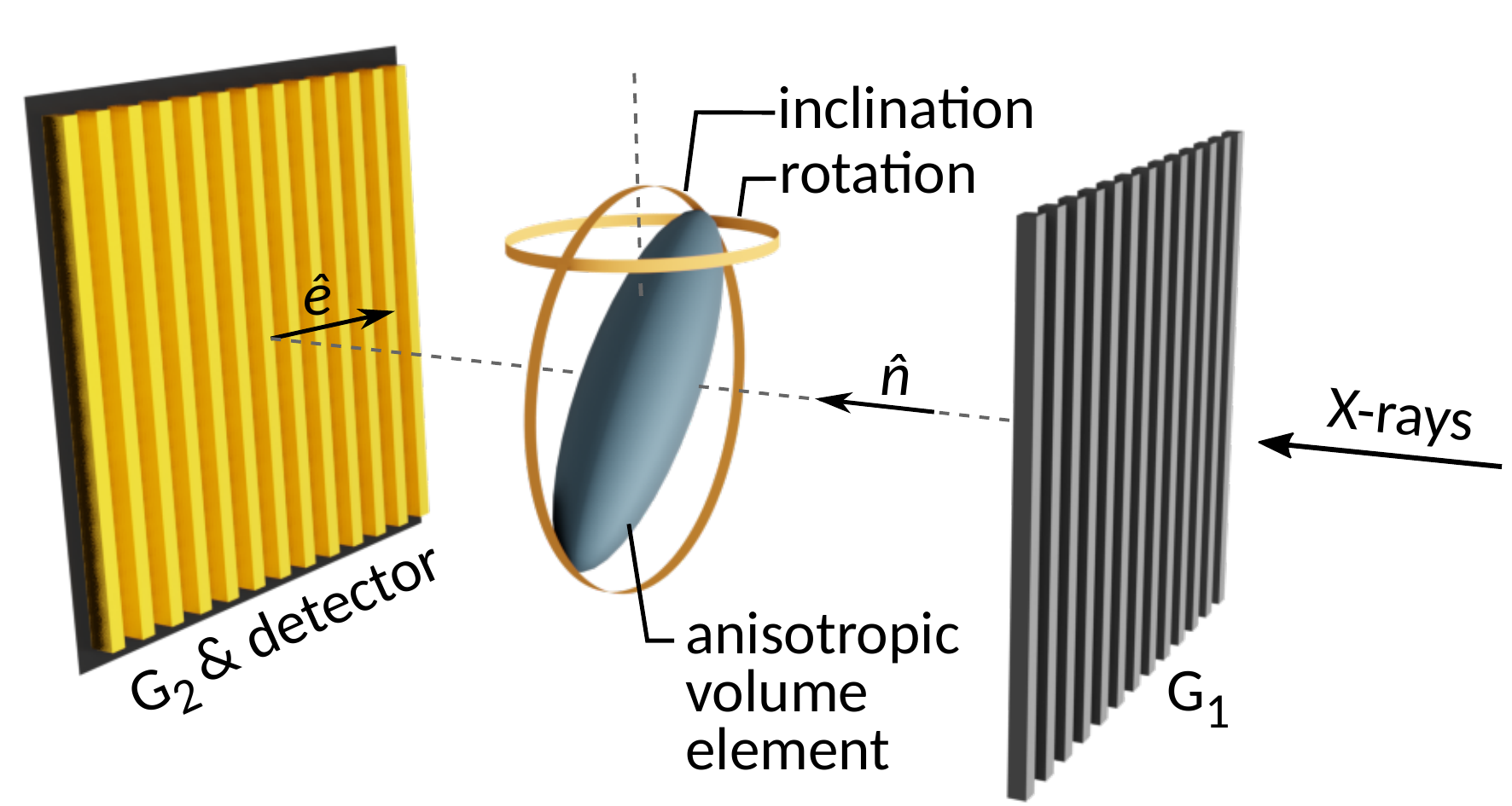}
\par\end{centering}
\caption{\label{fig:experiment-sktech}Sketch of a Talbot interferometer imaging
an anisotropic volume element. From left to right: Detector pixel
with analyzer grating G2, volume element with indicators for the rotations
and inclinations referred to in Fig.~\ref{fig:df-anisotropy}, and
the modulating grating G1. The optical axis and orientation of interferometer
sensitivity are indicated by $\hat{n}$ and $\hat{e}$ respectively.
(Figure adapted from \cite{Graetz2020})}
\end{figure}

\subsection{Linear tensor models and iterative reconstruction\label{subsec:lin-tens-recon}}

\begin{figure}
\begin{centering}
\includegraphics[viewport=0bp 0bp 490bp 324bp,width=0.8\textwidth]{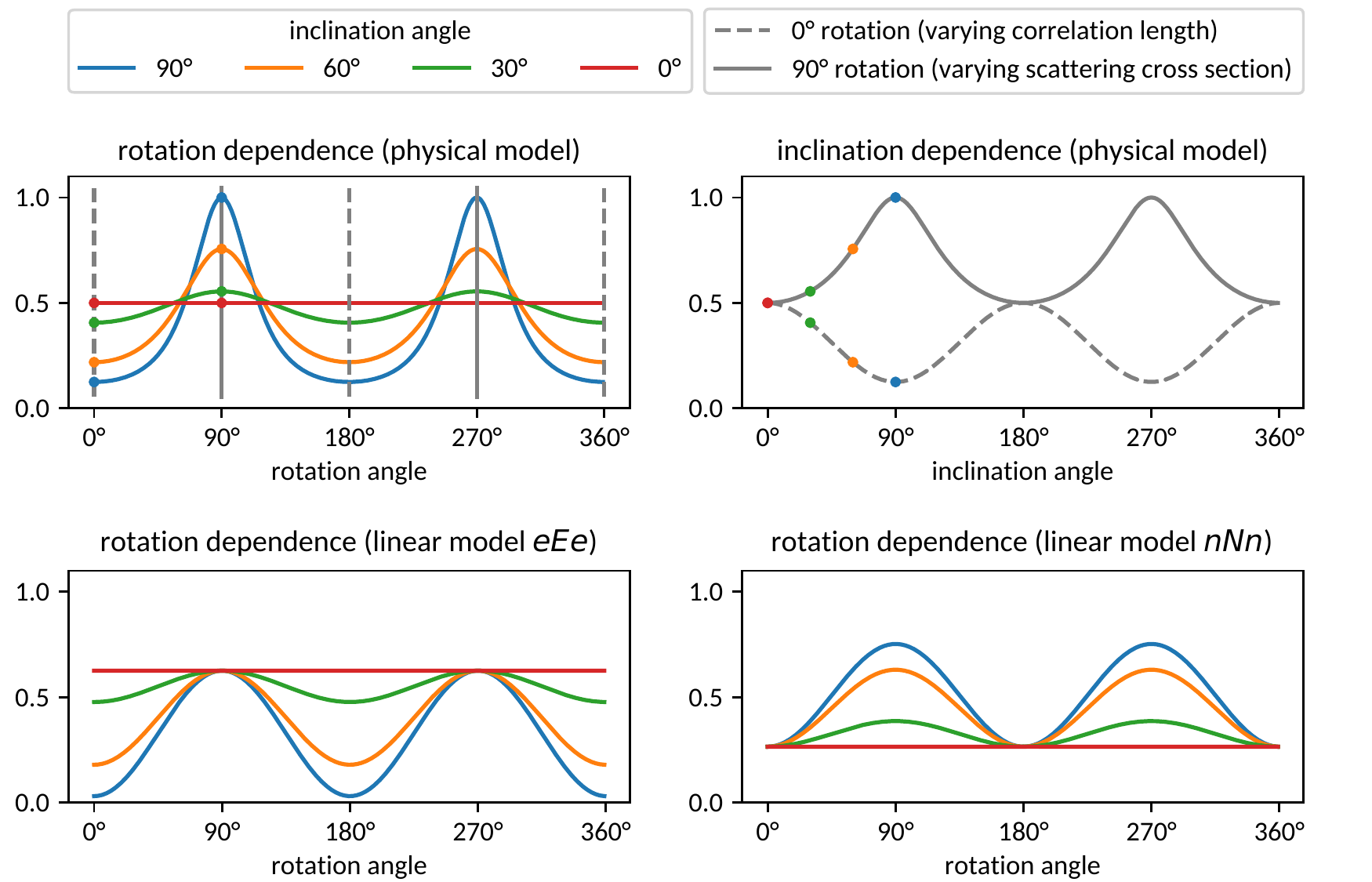}
\par\end{centering}
\caption{\label{fig:df-anisotropy}The different orientation dependencies of
the physical model of dark-field anisotropy and the linear tensor
models considered for reconstruction. Modeled is an elongated object
inclined with respect to a rotational axis perpendicular to both the
optical axis and the axis of interferometer sensitivity, cf.\ Fig.~\ref{fig:experiment-sktech}.
The original signal (upper left) is generated according to Eq.~\ref{eq:dark-field}
(with $\sigma_{3}/\sigma_{1}=\sigma_{3}/\sigma_{2}=2$) and arbitrarily
scaled to 1. The respective scales and offsets of the linear approximations
(bottom row) emerge from the reconstruction procedure.}
\end{figure}
\begin{figure}

\begin{centering}
\includegraphics[width=0.9\textwidth]{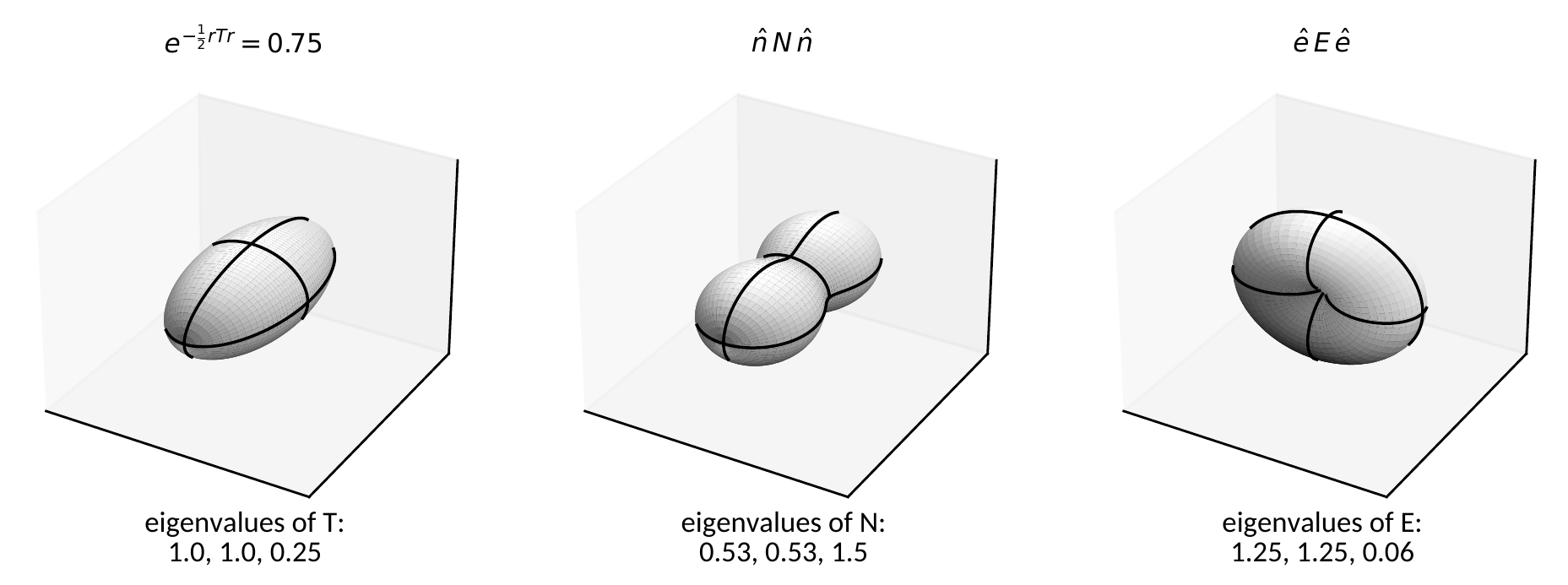}
\par\end{centering}
\caption{\label{fig:tensor-illustration}Illustrations of the considered tensors
and signal models: On the left, an anisotropic Gaussian mass density
distribution described by $\boldsymbol{T}$ (with $\sigma_{3}/\sigma_{1}=\sigma_{3}/\sigma_{2}=2$)
is represented by an iso-surface. The resulting dark-field signal
cannot be represented in three dimensions given its dependence on
two orientation vectors. On the center and right, the reconstructed
signal approximations parametrized by either of the two orientations
are shown. Both approximate models align with the principal orientation
of the original mass distribution. Systematic comparisons of eigenvectors
and eigenvalues are shown in Figs.~\ref{fig:errors}--\ref{fig:threetraj-props}.}

\end{figure}
A fundamental assumption for volume reconstruction from projections
is the linearity of the projection process. I.e., the sum of projections
of individual volume elements is expected to be equivalent to the
projection of the sum of the respective volume elements. In classic
X-ray tomography, this is ensured both by Beer's law of attenuation
and the assumption of an isotropic contrast mechanism. Eq.~\ref{eq:dark-field}
however is clearly non-linear with respect to the tensor $\boldsymbol{T}$
characterizing individual volume elements, and should, with regard
to tomographic volume reconstruction, be replaced by a suitable linear
surrogate.

Classic tensor tomography considers models of the form $\hat{r}\,\boldsymbol{U}\hat{r}$
with $\boldsymbol{U}=\boldsymbol{U}^{T}$, which are generally linear
(i.e., $\sum_{j}\hat{r}\,\boldsymbol{U}^{(j)}\hat{r}=\hat{r}(\sum_{j}\boldsymbol{U}^{(j)})\hat{r}\,$).
Given the actual dark-field signal dependence on both the optical
axis $\hat{n}$ and the direction $\hat{\boldsymbol{e}}$ of grating
sensitivity, two obvious models thus suggest themselves:
\begin{equation}
\mu_{\mathrm{DF}}(\hat{e},\hat{n},\boldsymbol{N})\approx\hat{n}\boldsymbol{N}\hat{n}\label{eq:nNn}
\end{equation}
and
\begin{equation}
\mu_{\mathrm{DF}}(\hat{e},\hat{n},\boldsymbol{E})\approx\hat{e}\boldsymbol{E}\hat{e}\,,\label{eq:eEe}
\end{equation}
which capture either of the dark-field contrast's orientation dependencies
respectively. The designations $\boldsymbol{N}$ and $\boldsymbol{E}$
for the respective symmetric $3\times3$ tensors have been chosen
for clear discernibility of both models. They both do represent gross
simplifications whose adequacy remains to be justified, and it therefore
is the aim of the present article to discuss their relation to the
actual contrast mechanism of dark-field imaging better described by
Eq.~\ref{eq:dark-field}.

To this end, tensors $\boldsymbol{N}$ and $\boldsymbol{E}$ will
be reconstructed from a set of scalar dark-field projections $\mu_{\mathrm{DF}}^{(i)}$
(enumerated by $i$) synthesized with Eq.~\ref{eq:dark-field} from
a given mass distribution tensor $\boldsymbol{T}$ and acquisition
geometry (characterized by a set of acquisition system orientations
$\hat{n}^{(i)},\hat{e}^{(i)}$), which will be detailed in Section~\ref{subsec:projection-geometry}.
The following iterative scheme as proposed in \cite{Dittmann2017} will be used for the reconstruction
of $\boldsymbol{N}$ and $\boldsymbol{E}$ from projection data respectively:
\begin{align}
U_{mn}^{(k)} & =U_{mn}^{(k-1)}+\lambda_{k}\overbrace{r_{m}^{(i_{k})}r_{n}^{(i_{k})}\smash[b]{(\underbrace{\mu_{\mathrm{DF}}^{(i_{k})}-\sum_{mn}r_{m}^{(i_{k})}U_{mn}^{(k-1)}r_{n}^{(i_{k})}}_{\text{residual}})}}^{\text{back projection}}\label{eq:tensor-recon}\\
\text{with}\quad U_{mn}^{(0)} & =0\nonumber \\
\text{and}\quad\hphantom{U_{mo}^{(0)}}\mathllap{\lambda_{k}} & =\lambda_{0}2^{-k/\tau},\quad\lambda_{0}\in\:]0,1]\nonumber \\
\text{assuming}\quad\hphantom{U_{mo}^{(0)}}\mathllap{\bigl\Vert\hat{r}\,\bigr\Vert} & =1,\nonumber 
\end{align}
with $k$ enumerating iterations, $i_{k}$ denoting the particular
projection $i$ considered at iteration $k$, and $\lambda_{k}$ being
a relaxation factor damping convergence. The specific back projection
procedure within Eq.~\ref{eq:tensor-recon} can be easily verified
to have two important properties: changes to $U_{mn}^{(k)}$ will
scale with the residual error between data and model, and the update
is performed by means of a pseudo-inverse of the linear tensor model
$\hat{r}\,\boldsymbol{U}\hat{r}$ such that immediate consistency
of $\hat{r}\,\boldsymbol{U}^{(k)}\hat{r}$ with the considered projection
$\mu_{\mathrm{DF}}^{(i_{k})}$ is achieved for $\lambda_{k}=1$. The
continuous reduction of $\lambda_{k}$ accounts for the expected inconsistency
between $\mu_{\mathrm{DF}}$ and the linear tensor models by gradually
reducing the impact of individual projections $\mu_{\mathrm{DF}}^{(i)}$
and thus their specific order of consideration as convergence to a
mean solution progresses.

In total, each individual projection will here be considered 25 times
throughout the iterative process, with $\lambda_{0}=.2$, $\tau=3k_{\max}/25$,
and $k_{\max}$ being 25 times the number of total projections $\mu_{\mathrm{DF}}^{(i)}$
(cf.\ following Section). 

\subsection{Projection geometry\label{subsec:projection-geometry}}

\begin{figure}
\begin{centering}
\includegraphics[viewport=0bp 9.5bp 195bp 165bp,clip,width=0.3\textwidth]{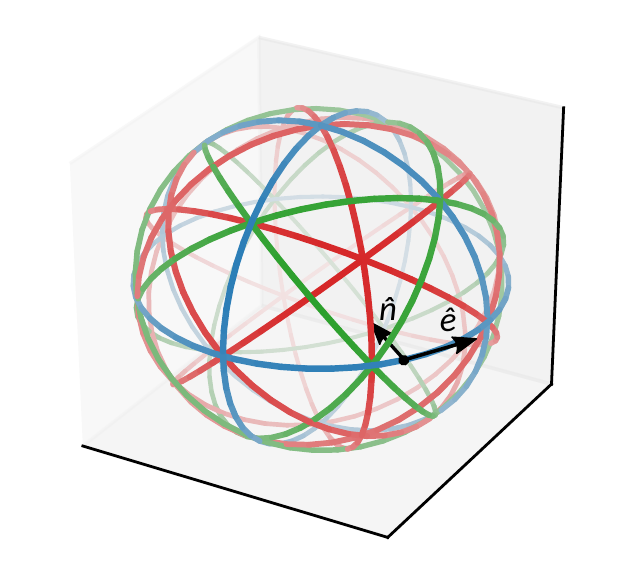}
\par\end{centering}
\caption{\label{fig:acq-geom}Circular scanning trajectories (expressed in
the sample coordinate system) about a volume element in the center
of the circumscribed sphere. Each trajectory describes a set of projection
directions $\hat{n}$ pointing towards the center and associated tangential
orientations of interferometer sensitivity $\hat{e}$. For better
visualization, they have been grouped into trajectories about the
coordinate axes (blue), about their face diagonals (red) and about
their space diagonals (green). Intersections of trajectories represent
points of constant projection direction $\hat{n}$ at varying dark-field
sensing directions $\hat{e}$. For the present study, each trajectory
is sampled at 29 equidistant points, yielding a total of 377 distinct
combinations $(\hat{n}^{(i)},\text{\ensuremath{\hat{e}}}^{(i)})$
enumerated by $i$.}
\end{figure}
The reconstruction of individual tensor voxels from scalar projections
requires a number of such projections sufficiently covering the parameter
space of the signal model, i.e., sufficiently covering variations
in both orientations $\hat{n}$ and $\hat{e}$. Figure~\ref{fig:acq-geom}
depicts circular trajectories about a volume element in the origin.
Each point on the trajectories represents a different orientation
of the optical axis or direction of projection $\hat{n}$ pointing
towards the center, and the respective orientation $\hat{e}$ of grating
sensitivity is here defined tangential to the respective circular
rotation orbits. Each trajectory is normal to one of the coordinate
axes (blue) or to one of their diagonals (red and green), yielding
a total of 13. Such a geometry is, in practice, easily realized by
placing a sample in different orientations on the rotary stage of
a typical tomography setup. With regard to the present simulations
of individual volume elements, each trajectory is sampled at 29 equidistant
points, i.e., a total of 337 scalar dark-field values at numerous
combinations of $\hat{n}$ and $\hat{e}$ is synthesized for each
simulated volume element.

\subsection{Parameter exploration}

The parameter space of the physical dark-field model given in Eq.~\ref{eq:dark-field}
consists of the to-be-reconstructed tensor characterizing the volume
element on the one hand and the parameters $\hat{n}$ and $\hat{e}$
of the considered projection on the other hand. While the coverage
of projection parameters, which correspond to the imaging procedure,
has just been discussed, an adequate exploration scheme for the space
of anisotropic volume elements described by $\boldsymbol{T}$ shall
be addressed here.

At first it can be easily confirmed that the absolute scale of $\boldsymbol{T}$
does not affect the orientation dependence of $\mu_{\mathrm{DF}}$.
It is therefore sufficient, without loss of generality, to require
$\text{trace}(\boldsymbol{T})=1=\sum_{i}\sigma_{i}^{-2}$ for the
present purposes, thereby implicitly constraining one eigenvalue and
reducing the parameter space to the remaining two. Further, permutations
of equivalent sets of eigenvalues correspond to rotations, which are
treated separately. Based on these considerations, the following definitions
can be made with regard to an exhaustive coverage of the parameter
space:
\begin{equation}
\begin{alignedat}{1}\sigma_{1}^{-2} & \in[0,1/3]\\
\sigma_{2}^{-2} & \in[0,1/2]\\
\sigma_{3}^{-2} & \in[1/3,1]\\
\text{with}\quad\hphantom{\sum_{i}\sigma_{i}^{-2}}\mathllap{\sigma_{1}^{-2}} & \leq\sigma_{2}^{-2}\leq\sigma_{3}^{-2}\\
\text{and}\quad\sum_{i}\sigma_{i}^{-2} & =1\:.
\end{alignedat}
\label{eq:eigenvalue-ranges}
\end{equation}

\begin{figure}

\centering{}\includegraphics[width=0.9\textwidth]{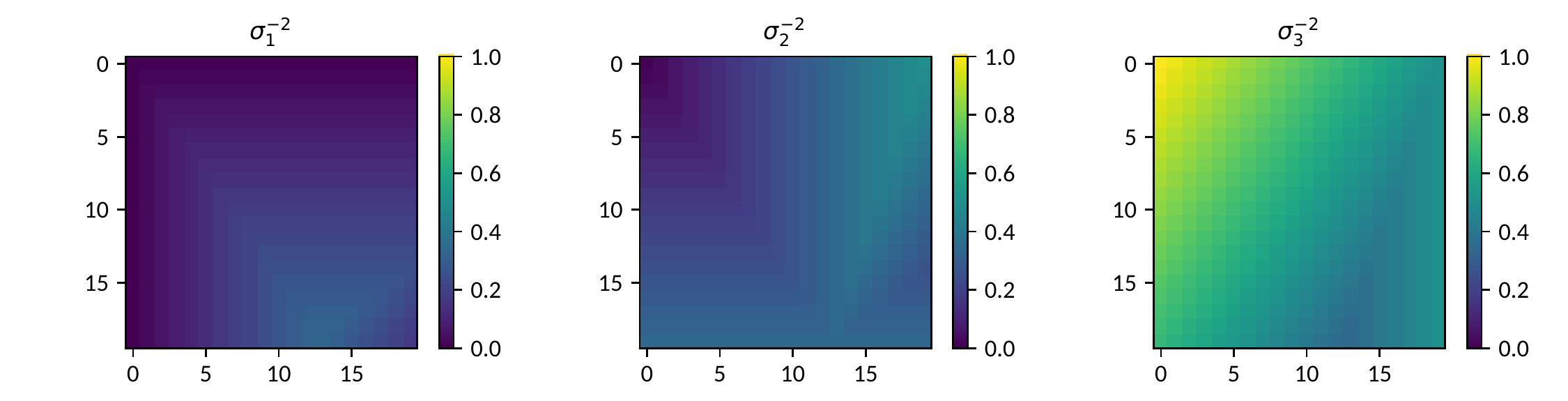}\caption{\label{fig:eigenvalue-ranges}Coverage of the parameter space (on
a 20${}\times{}$20 grid) of possible eigenvalue combinations in $\boldsymbol{T}$
according to Eq.~\ref{eq:eigenvalue-ranges}. A tolerated amount
of redundancy can be observed in the bottom right region.}
\end{figure}
The eigenvalues of $\boldsymbol{T}$ parametrizing the synthesization
model are sampled on a regular grid covering $20\times20$ combinations
of $\sigma_{1}^{-2}$ and $\sigma_{2}^{-2}$ within the specified
ranges, with $\sigma_{3}^{-2}$ resulting from the given constraints.
The particular coverage shown in Figure~\ref{fig:eigenvalue-ranges}
results from the generation of two orthogonally oriented linear ramps
ranging 0 to 1/3 and from 0 to 1/2 respectively, which are then complemented
by a third map according to the normalization requirement. After sorting
the generated eigenvalues, Fig.~\ref{fig:eigenvalue-ranges} emerges.

\section{Results}

For each feasible triplet of eigenvalues, 300 randomly oriented instances
of a volume element described by $\boldsymbol{T}$ are generated (by
means of random rotation matrices constructed according to \cite{Arvo1992}),
yielding 120$\,$000 instances in total. Corresponding tensors $\boldsymbol{N}$
and $\boldsymbol{E}$ are reconstructed according to Section~\ref{subsec:lin-tens-recon}
from the volume element's dark-field signals synthesized using Eq.~\ref{eq:dark-field}
based on its mass distribution tensor $\boldsymbol{T}$ and the given
projection geometry (cf.\ Fig.~\ref{fig:acq-geom}). Illustrating
examples are given in Figures \ref{fig:df-anisotropy} and \ref{fig:tensor-illustration}.
The relation between the simplified and the complete model of dark-field
anisotropy is assessed by comparing their tensors' eigenvectors and
-values.

\subsection{Goodness of fit and reproduction of dominant orientations}

\begin{figure}
\begin{centering}
\includegraphics[width=0.7\textwidth]{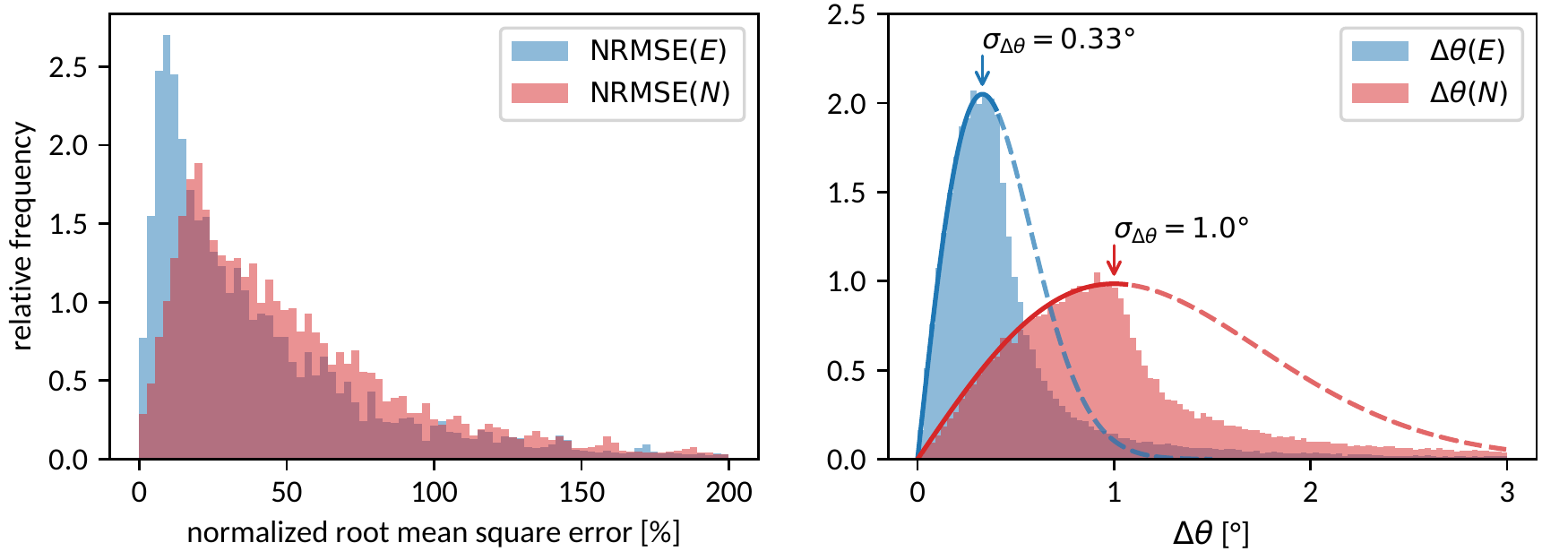}
\par\end{centering}
\caption{\label{fig:errors}Histograms showing the normalized root mean square
errors (NRMSE, cf.\ Eq.~\ref{eq:nrmse}) and angular deviations
(Eq.~\ref{eq:angle-dev}) of the fitted approximative tensor models
with respect to the synthesized dark-field data and mass distribution
tensors. The angular deviations are compared to (unnormalized) Gaussian
distributions of inclination angles (integrated over the azimuthal
degree of freedom, cf.\ Eq.~\ref{eq:gaussian-pdf}). These apparently
describe the empiric distributions up to their maximum and are used
for quantification purposes. }
\end{figure}

Figure~\ref{fig:errors} (left) shows normalized root mean square
errors 
\begin{equation}
\text{NRMSE}=\frac{\sqrt{\overline{(\mu_{\mathrm{DF}}^{(i)}-\hat{r}^{(i)}\boldsymbol{U}\hat{r}^{(i)})^{2}}}}{\overline{\mu_{\mathrm{DF}}^{(i)}}}\label{eq:nrmse}
\end{equation}
of the linear tensor models (\ref{eq:nNn}) and (\ref{eq:eEe}) with
respect to the noiseless input data $\mu_{\mathrm{DF}}^{(i)}$. The
statistical distribution arises both from the range of considered
anisotropies and orientations. While the distributions peak at 10-20\%
NRMSE, values over 100\% are still encountered. These deviations reflect
the approximative nature of the linear tensor models.

Nevertheless, the principal orientation, which is one of the central
concerns of tensor tomography, is reproduced to a typical precision
of $0.33{^\circ}$ and $1.0{^\circ}$ respectively for the considered
reconstruction models: Figure~\ref{fig:errors} (right) compares
the eigenvectors corresponding to either the smallest (in the case
of $\boldsymbol{E}$) or largest (in the case of $\boldsymbol{N}$)
eigenvalues to the original synthesized orientation of $\boldsymbol{T}$
(cf.\ Figure~\ref{fig:tensor-illustration} for a visualization
of the respective tensors and their extents). The considered eigenvector
of each tensor indicating a volume element's principal orientation
is denoted $\hat{v}_{T}$, $\hat{v}_{E}$ and $\hat{v}_{N}$ respectively,
whereby $\hat{v}_{T}$ represents the ground truth. The reconstruction
error (with respect to orientation) induced by the simplified models
is measured by the relative angles
\begin{align}
\Delta\theta(\boldsymbol{E}) & =\arccos|\hat{v}_{T}\cdot\mathrlap{\hat{v}_{E}}\hphantom{\hat{v}_{N}}|\nonumber \\
\text{and }\Delta\theta(\boldsymbol{N}) & =\arccos|\hat{v}_{T}\cdot\hat{v}_{N}|\:.\label{eq:angle-dev}
\end{align}
The observed orientation error distributions are compared to a normal
distribution of inclination angles, which, due to integration over
the azimuthal angle, takes the form
\begin{equation}
\mathrm{PDF}(\Delta\theta,\sigma_{\Delta\theta})\propto\sin(\Delta\theta)e^{-\frac{1}{2}\frac{\Delta\theta^{2}}{\sigma_{\Delta\theta}}}\approx\Delta\theta e^{-\frac{1}{2}\frac{\Delta\theta^{2}}{\sigma_{\Delta\theta}^{2}}}.\label{eq:gaussian-pdf}
\end{equation}
It can be easily shown to exhibit its maximum at $\Delta\theta=\sigma_{\Delta\theta}$
in the small angle approximation $\sin(\Delta\theta)\approx\Delta\theta$.
Figure~\ref{fig:errors} (right) shows to be consistent with this
distribution up to the maximum value, i.e., up to its standard deviation
parameter. 

\subsection{Eigenvalues}

\begin{figure}
\begin{centering}
\includegraphics[width=1\textwidth]{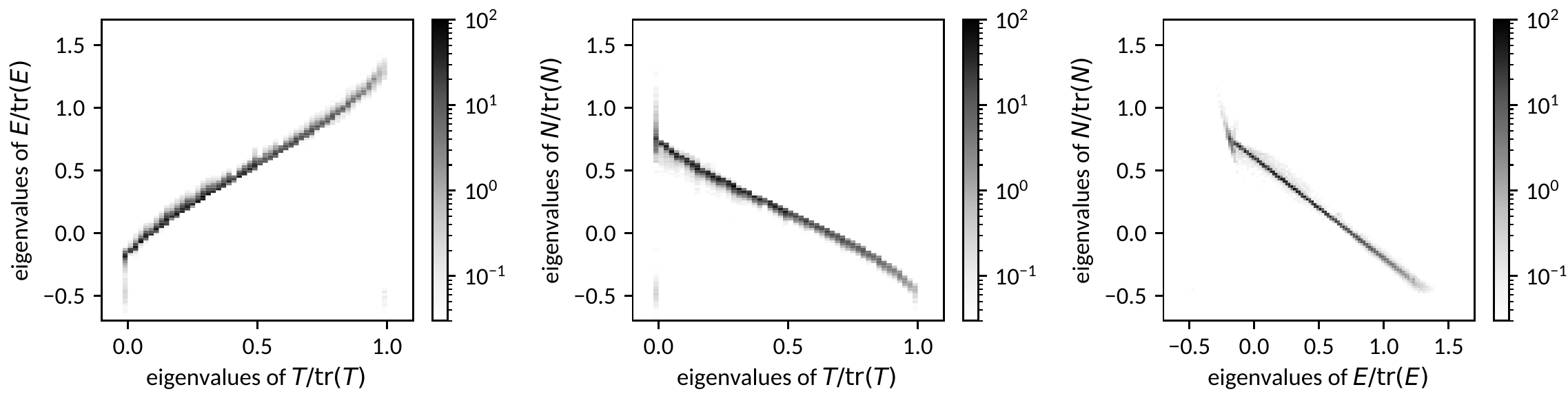}
\par\end{centering}
\centering{}\caption{\label{fig:eigenspectra-correlations}Pairwise correlations of the
eigenspectra of tensors describing dark-field anisotropy in different
ways (cf.\ Fig.~\ref{fig:tensor-illustration}). For each pair of
the tensors $\boldsymbol{T}$, $\boldsymbol{N}$ and $\boldsymbol{E}$
(cf.\ Eqs.~\ref{eq:dark-field}, \ref{eq:nNn} and \ref{eq:eEe}),
their respective normalized and sorted eigenvalues (in reverse order
for $\boldsymbol{N}$) are plotted against each other in 2D density
histograms. Instances of $\boldsymbol{N}$ and $\boldsymbol{E}$ have
to this end been reconstructed according to Section~\ref{subsec:lin-tens-recon}
from dark-field signals based on Eq.~\ref{eq:dark-field} and given
instances of $\boldsymbol{T}$, generated according to Fig.~\ref{fig:eigenvalue-ranges}
in combination with random rotation matrices. The eigenvalues of both
linear tensor models are found to be approximately linear in $\sigma_{i}^{-2}$
(eigenvalues of $\boldsymbol{T}$)}
 
\end{figure}

Figure~\ref{fig:eigenspectra-correlations} reveals fuzzy, yet almost
linear relations among the normalized and sorted eigenvalues of all
models despite the strong approximations involved. Due to the expected
anti-correlation of the eigenvalues of $\boldsymbol{N}$ with those
of $\boldsymbol{T}$ and $\boldsymbol{E}$ (cf.\ the illustrating
example given in Figure~\ref{fig:tensor-illustration}), they are
sorted in reverse order prior to comparison.

Model (\ref{eq:eEe}) parametrized by $\boldsymbol{E}$ exhibits an
almost proportional relation to the original signal generating mass
distribution tensor $\boldsymbol{T}$ with moderate deviation from
its positive definite nature. Model (\ref{eq:nNn}) parametrized by
$\boldsymbol{N}$ in contrast exhibits an inverse relation to $\boldsymbol{T}$
with the spectrum of eigenvalues being notably shifted towards negative
values. The mapping of eigenvalues nevertheless remains approximately
linear as opposed to an actual reciprocal relation. The direct comparison
of the normalized eigenspectra of $\boldsymbol{N}$ and $\boldsymbol{E}$
reveals -- on average -- a perfectly linear relation, indicating
that both capture highly similar information given the present acquisition
geometry (see Fig.~\ref{fig:acq-geom}). Deviations from the apparent
mean curves relating the different models reflect both statistical
variances among different orientations of the same volume element
as well as systematic deficiencies of the simplified models. As these
effects are not separable in practical applications, no further effort
is made to investigate their individual contributions.

The case of one or two eigenvalues of $\boldsymbol{T}$ being exactly
zero corresponds to the limit of infinitely extended volume elements,
causing an extreme orientation dependence and divergence of the dark-field
signal according to Eq.~\ref{eq:dark-field} and thus also unstable
(highly orientation dependent) reconstruction results in $\boldsymbol{N}$
and $\boldsymbol{E}$, as can be observed in Fig.~\ref{fig:eigenspectra-correlations}.
This limit is expected to be only of academic relevance.

\subsection{Reconstruction from three orthogonal trajectories}

\begin{figure}
\begin{centering}
\includegraphics[width=0.7\textwidth]{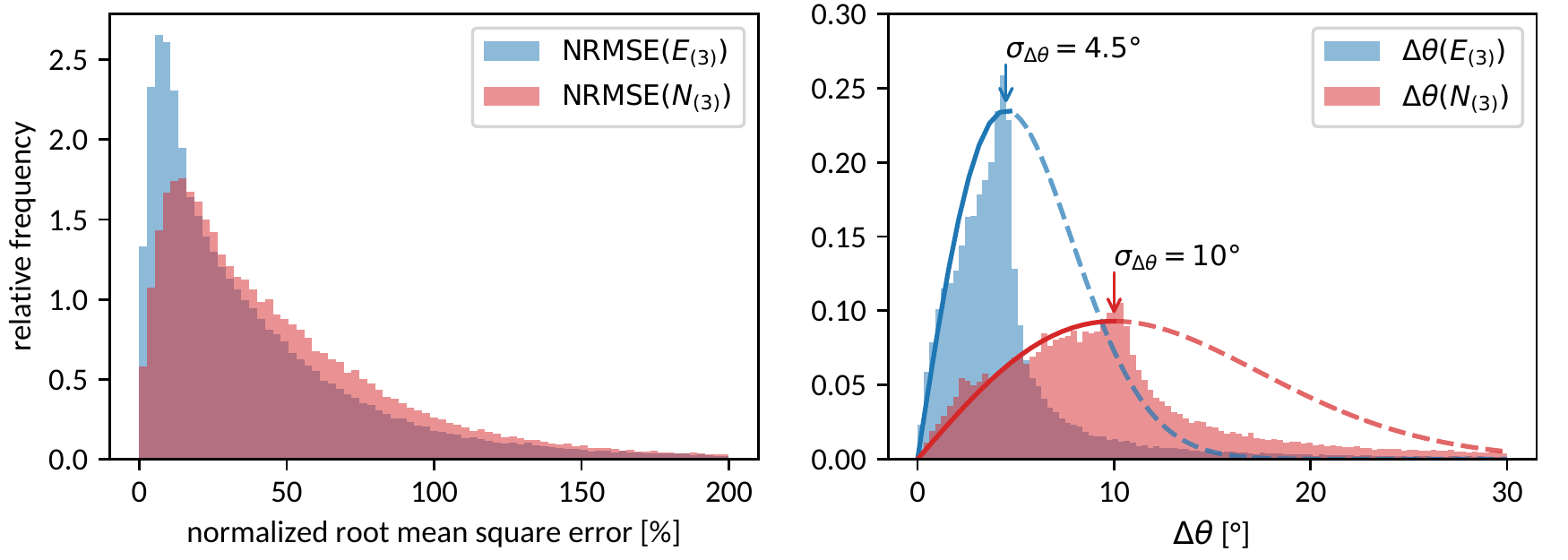}
\par\end{centering}
\begin{centering}
~
\par\end{centering}
\begin{centering}
\includegraphics[width=1\textwidth]{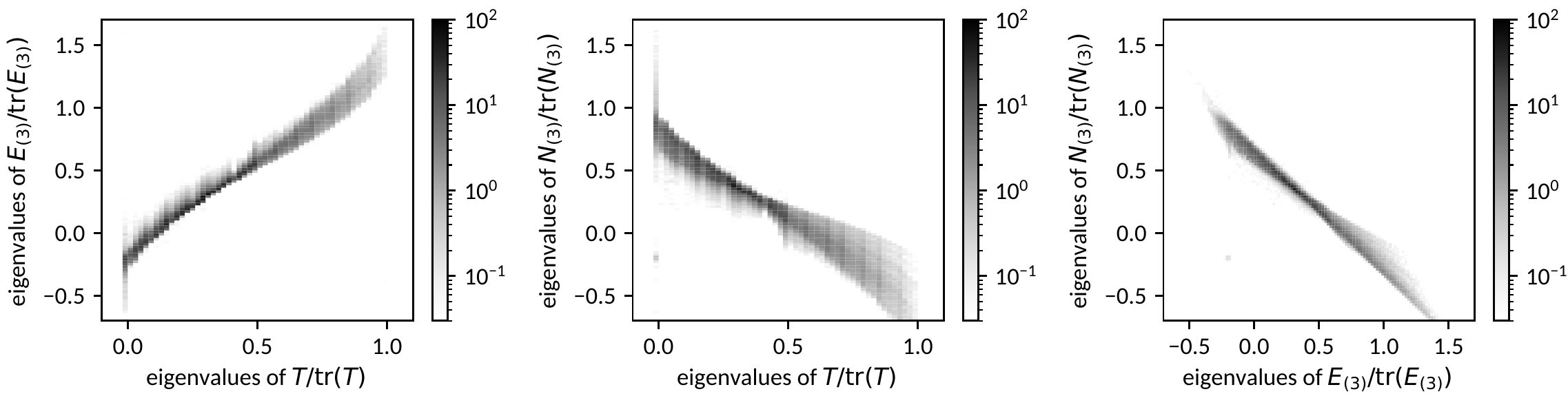}
\par\end{centering}
\caption{\label{fig:threetraj-props}Comparison of NRMSEs, principal orientations
and eigenspectra analogous to Figures~\ref{fig:errors}--\ref{fig:eigenspectra-correlations}
when reconstructing tensors $\boldsymbol{N}$ and $\boldsymbol{E}$
from a reduced set of dark-field projections acquired along only three
orthogonal projection trajectories (Figure~\ref{fig:acq-geom}, indicated
in blue).}

\end{figure}

If dark-field anisotropy was actually adequately described by either
of the linear models (\ref{eq:nNn}) or (\ref{eq:eEe}), three orthogonal
projection trajectories (Fig.~\ref{fig:acq-geom}, blue) would fundamentally
be sufficient to fully determine the respective tensor (also in a
tomographic setting, cf.\ e.g.\ \cite{Defrise2005}). As such a
reduced set of trajectories is highly desirable with regard to practical
data acquisition, it shall therefore be briefly considered here as
well. Figure~\ref{fig:threetraj-props} depicts, analog to Figs.~\ref{fig:errors}--\ref{fig:eigenspectra-correlations},
the relations between the considered models' tensors. Despite the
reduced set of data, which generally gives reason to expect fewer
model inconsistencies, the observed distribution of root mean square
errors is still qualitatively comparable to that found previously
in Fig.~\ref{fig:errors}. And although the visibly diffused relations
between the tensors' eigenspectra reflect a notably degraded relation
between the reconstructed tensors and the original input, principal
orientations are still roughly reproduced within an error margin of
about 4.5° to 10°. 

\section{Discussion }

While several proofs of concept plausibly demonstrating tomographic
reconstruction of sub-resolution anisotropy based on various heuristic
signal models were given in previous literature, explicit validations
have been lacking so far. A central assumption explicitly or implicitly
shared by all current approaches is that 3D dark-field anisotropy
can be described as a function of a single orientation vector. This
has been, deriving from planar dark-field anisotropy, the interferometer's
direction of sensitivity. A detailed discussion of dark-field origination
given recently \cite{Graetz2020} concludes that a complete description
of general dark-field anisotropy for arbitrarily oriented structures
further exhibits a non-negligible dependence also on the relative
orientation of the optical axis (the direction of projection). I.e.,
dark-field contrast will likewise vary for anisotropic structures
rotating about the axis of grating sensitivity (changing their inclination
with respect to the optical axis) as for the classical case of objects
rotating about the optical axis (changing their orientation with respect
to the interferometer's gratings). General dark-field anisotropy is
thus fundamentally a function of two orientations: the axis of interferometer
sensitivity, and the axis of projection.

With regard to tensor tomography, this observation has two remarkable
implications: Foremost, it obviously raises the question why the present
approaches to anisotropic dark-field tomography do nevertheless produce
plausible results. And as neglecting parts of the orientation dependence
is apparently tolerable, this also indicates that dark-field anisotropy
may likewise be approximable as a function of the optical axis, modeling
variations in scattering cross section as opposed to variations in
auto-correlation properties. As both effects have opposite relations
with the considered structure's extents (cf.\ Fig.~\ref{fig:df-anisotropy}
upper right), the net relation of either model to the signal generating
structure is non-obvious. The purpose of the present study therefore
was to systematically investigate the actual relations based on the
physically motivated signal model derived and verified previously
(\cite{Graetz2020}) based on the current state of knowledge on dark-field
origination.

As expected, the approximative models can exhibit large root mean
square errors with regard to the actual signal. When providing input
data decently covering both orientation dependencies, both linear
models nevertheless manage to reproduce the principal orientation
of the original mass distribution tensor up to a typical accuracy
of 0.33° to 1°. The normalized eigenvalues, which do encode the actual
aspect ratios of anisotropic volume elements, are thereby found to
be roughly linear in the normalized inverse variances of the original
mass distribution (as opposed to being linear e.g.\ in its extents)
for both models. As this is the scaling behavior expected along the
axis of interferometer sensitivity, it is consistent with the observation
that model (\ref{eq:eEe}) yields slightly better reconstruction results.

Although the true complexity of dark-field anisotropy generally mandates
a rather extensive data acquisition scheme, the consequences of using
only a minimal set of three circular acquisition trajectories has
been explicitly considered as well. While the quantitative relation
between input and reconstruction is, as expected, notably degraded,
principal orientations for isolated volume elements could still be
roughly recovered to 5° to 10° accuracy. As it is conjecturable that
the additional influences of noise and signal superpositions (in the
tomography use case) will further challenge the stability of these
reconstructions in practical applications, more comprehensive acquisition
schemes as considered initially are highly recommended though.

\section{Conclusion}

The practical feasibility of dark-field tensor volume tomography depends
on the applicability of linear approximations to actual dark-field
signal anisotropy, given that highly non-linear dependencies can void
the ability to solve the problem of volume reconstruction from projections.
Given a previously validated dark-field signal model, an exhaustive
exploration of the possible parameter space for anisotropic volume
elements has been performed in order to establish the general ability
of approximative linear tensor models of dark-field anisotropy to
recover central properties of the considered volume elements. While
the reconstruction of orientation and aspect ratios is fundamentally
subject to statistical variation already in the absence of signal
noise, the theoretically achievable precision of 1° and better is
highly encouraging with regard to quantitative tensor valued volume
tomography. Although linear tensor models (in 3D space) would generally
be fully determined with data from only three orthogonal projection
trajectories, the data acquisition scheme must however nevertheless
be guided by the true complexity of the actual dark-field anisotropy. 

\bibliographystyle{plain}

\end{document}